\documentclass[twocolumn,prl]{revtex4}
\usepackage{graphics,graphicx}
\begin{document}
\title{Spin Correlations in the Geometrically Frustrated Pyrochlore Tb$_{2}$Mo$_{2}$O$_{7}$}
\author{D.K.~Singh$^{1}$}
\author{J.S.~Helton$^{1}$}
\author{S.~Chu$^{2}$}
\author{T.H.~Han$^{1}$}
\author{C.J.~Bonnoit$^{1}$}
\author{S.~Chang$^{3,4}$}
\author{H.J.~Kang$^{3,4}$}
\author{J.W.~Lynn$^{3}$}
\author{Y.S.~Lee$^{1}$}
\affiliation{$^{1}$Department of Physics, Massachusetts Institute
 of Technology, Cambridge, MA 02139, USA}
\affiliation{$^{2}$Center for Materials Science and Engineering,
Massachusetts Institute of Technology, Cambridge, MA 02139, USA}
\affiliation{$^{3}$NIST Center for Neutron Research, Gaithersburg,
MD 20899, USA} \affiliation{$^{4}$Department of Materials Science
and Engineering, University of Maryland, College Park, Maryland
20742, USA}

\begin{abstract}
We report neutron scattering studies of the spin correlations of the
geometrically frustrated pyrochlore Tb$_{2}$Mo$_{2}$O$_{7}$ using
single crystal samples.  This material undergoes a spin-freezing
transition below $T_g \simeq 24$~K, similar to
Y$_{2}$Mo$_{2}$O$_{7}$, and has little apparent chemical disorder.
Diffuse elastic peaks are observed at low temperatures, indicating
short-range ordering of the Tb moments in an arrangement where the
Tb moments are slightly rotated from the preferred directions of the
spin ice structure. In addition, a $\vec{Q}$-independent signal is
observed which likely originates from frozen, but completely
uncorrelated, Tb moments. Inelastic measurements show the absence of
sharp peaks due to crystal field excitations. These data show how
the physics of the Tb sublattice responds to the glassy behavior of
the Mo sublattice with the associated effects of lattice disorder.
\end{abstract}

 \maketitle

Geometrically frustrated magnets can exhibit unusual phenomena at
low temperatures, such as spin glass\cite{Greedan,Gardner_PRL}, spin
ice\cite{Ramirez,Harris_ice,Gingras_science}, and spin
liquid\cite{Gardner_SL,Helton} states.  The frustrated pyrochlore
compound Tb$_{2}$Mo$_{2}$O$_{7}$ stands at an interesting
crossroads, being a representative of both the Mo-based family
$R_{2}$Mo$_{2}$O$_{7}$ ($R={\rm rare~earth}$) and other Tb-based
compounds Tb$_{2}M_{2}$O$_{7}$ ($M={\rm metal}$).  As a function of
the $R$-site radius, the $R_2$Mo$_2$O$_7$ compounds exhibit a
metal-insulator transition between ferromagnetic metal states to
spin glass insulators.\cite{Hanasaki_PRL,Kezsmarki_PRB} The
ferromagnetic metals (such as Nd$_{2}$Mo$_{2}$O$_{7}$ and
Sm$_{2}$Mo$_{2}$O$_{7}$) show an interesting interplay between the
spin configuration and the transport properties, which can give rise
to anomalous Hall and Nernst
signals.\cite{Tokura_science,Tokura_nernst} The spin glass
insulators (such as Y$_{2}$Mo$_{2}$O$_{7}$ and
Tb$_{2}$Mo$_{2}$O$_{7}$) have generated much interest due to the
apparent lack of chemical disorder and the relation to geometrical
frustration.\cite{Greedan,Gardner_PRL} Outside of the Mo-family,
pyrochlore compounds with Tb as the rare earth have been extensively
studied because the combination of exchange and dipolar interactions
coupled with moderate spin anisotropy can lead to novel
ground-states depending on the metal ion.  For example,
Tb$_{2}$Ti$_{2}$O$_{7}$ has a spin liquid ground
state\cite{Gardner_SL}, while Tb$_{2}$Sn$_{2}$O$_{7}$ exhibits
magnetic order\cite{Mirebeau_Sn}, and both show persistent spin
dynamics down to the lowest measured
temperatures.\cite{Gardner_SL,Dalmas_PRL,Bert_PRL}

Tb$_{2}$Mo$_{2}$O$_{7}$ crystallizes in the $Fd\bar{3}m$ cubic space
group, and both the Tb and Mo sublattices form three dimensional
networks of corner sharing tetrahedra. Spin-freezing is observed
below $T_g \simeq 24$~K, and diffuse scattering from the Tb moments
has been observed with neutron scattering on powder
samples.\cite{Greedan,Gaulin_PRL,Apetrei_PRL} Remarkably, the spins
keep fluctuating down to the lowest temperatures measured in $\mu$SR
studies\cite{Dunsiger}. Currently, the spin structure of the
short-range order is not known, nor has there been a precise
determination of the correlation lengths.  Also, not much is known
about the possible effects of disorder, if present.  In this paper,
we present neutron scattering measurements on a single crystal
sample which answer some of these questions and also uncover new
features related to the interaction between the Tb and Mo
sublattices.

Single crystal samples of Tb$_{2}$Mo$_{2}$O$_{7}$ were grown using
the floating-zone technique in an image furnace.  A mixture of
Tb$_{4}$O$_{7}$ (99.97$\%$) and MoO$_{2}$ (99.99$\%$) was thoroughly
ground and pressed into feed rods, and the crystal growth was
carried out in an Ar atmosphere. The samples were confirmed to be
single phase Tb$_{2}$Mo$_{2}$O$_{7}$ by x-ray and neutron
diffraction.  The magnetic susceptibility, shown in the inset of
Fig. 1(a), is consistent with previous results on powders. The
difference between the zero-field cooled (ZFC) and field cooled (FC)
susceptibility indicate a spin-freezing temperature of $T_g \simeq
24$~K.  Fitting the high temperature susceptibility data ($T>100$~K)
to a Curie-Weiss law yields a Curie-Weiss temperature of
$\Theta_{CW}$ $\simeq 12$~K and an effective moment of
$\sim9.5~\mu$$_{B}$. The size of the moment is consistent with large
Tb$^{3+}$ moments\cite{Gingras_PRB} dominating the response. The
nature of the magnetic couplings is not so clear; currently, the
values of the Tb-Tb, Tb-Mo, and Mo-Mo interactions in
Tb$_{2}$Mo$_{2}$O$_{7}$ are not known. In Y$_{2}$Mo$_{2}$O$_{7}$
(where Y is non-magnetic), the Curie-Weiss temperature of about
$-200$~K suggests that the Mo-Mo interaction is strongly
antiferromagnetic.\cite{Gardner_PRL} Y$_{2}$Mo$_{2}$O$_{7}$ has a
spin-freezing temperature ($T_g \simeq 22.5$~K) similar to that in
Tb$_{2}$Mo$_{2}$O$_{7}$; hence, it is reasonable to assume that the
Mo sublattice is primarily responsible for the freezing transition.
Neutron measurements on Y$_{2}$Mo$_{2}$O$_{7}$ have shown that
dynamical short range order of the Mo sublattice develops below room
temperature.\cite{Gardner_PRL}  To better understand how the
Tb$^{3+}$ ions respond to the glassy behavior of the Mo sublattice,
we have carried out a series of neutron scattering measurements.

\begin{figure}
\centering
\includegraphics[width=8.2cm]{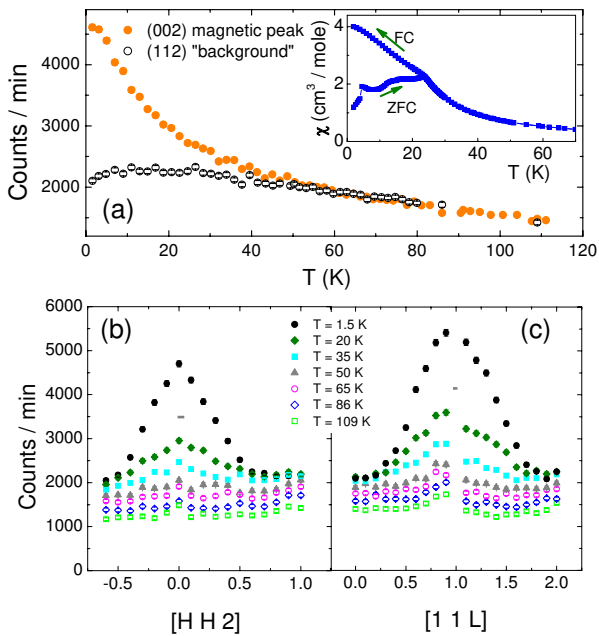} \vspace{-1mm}
\caption{(color online) Elastic scattering measured on SPINS
spectrometer.  (a) Intensity at the (002) diffuse magnetic peak
position and the ``background'' signal measured at the (112)
position as a function of temperature. Inset: Magnetic
susceptibility measured with $H=10$~Oe along the [111] direction
using a SQUID magnetometer. Scans through the (b) (002) position and
the (c) (111) position at various temperatures. The horizontal bar
indicates the instrumental resolution} \vspace{-4mm}
\end{figure}

The neutron experiments were performed using a 0.49 gram single
crystal of Tb$_{2}$Mo$_{2}$O$_{7}$ at the NIST Center for Neutron
Research. The SPINS and BT9 triple axis spectrometers were used with
fixed final neutron energies of 4.5 meV and 14.7 meV, respectively.
At SPINS, we employed the horizontally focused analyzer with
collimator sequence $80^{'}$$-$sample$-$Be$-$radial$-$open. At BT9,
a flat analyzer was used with collimations
$40^{'}$$-$$48^{'}-$PG$-$sample$-$$40^{'}$$-$$90^{'}$. The crystal
was oriented in the ($HHL$)-zone and placed in a $^4$He cryostat.

On SPINS, we observed the onset of diffuse elastic peaks upon
cooling, and representative scans through the (002) and (111)
positions are shown in Fig. 1(b) and (c).  At $T=1.5$~K, the peak
widths are large, spanning a considerable fraction of the Brillioun
zone, indicating short-range order.  Since the magnitude of the Tb
moment should be about 5 times larger than the Mo moment, we
identify these peaks as arising from short-range order on the Tb
sublattice. The intensity at the (002) diffuse peak position is
plotted as a function of temperature in Fig. 1(a), and the
``background'' signal measured at (112) is also plotted.  The onset
of the diffuse peak intensity occurs around $T \sim 2\,T_g$ and
increases smoothly upon cooling through $T_g$. This is similar to
the temperature dependence of elastic scattering observed for
Y$_{2}$Mo$_{2}$O$_{7}$\cite{Gardner_PRL}, indicating that the Tb
correlations are coupled to the developing Mo correlations.

\begin{figure}
\centering
\includegraphics[width=8.0cm]{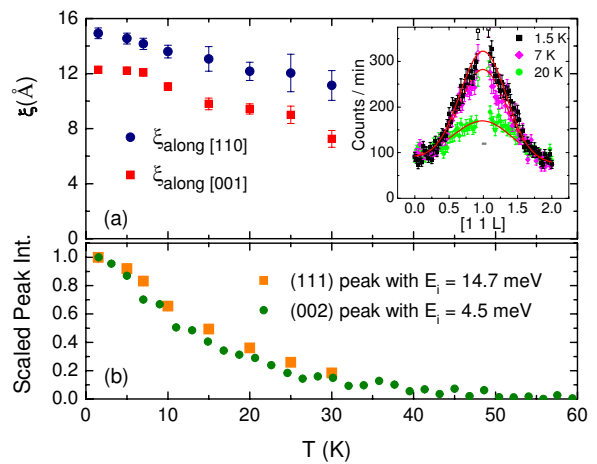} \vspace{-1mm}
\caption{(color online) (a) Correlation lengths of the short range
order along different symmetry directions as a function of
temperature. Inset: Representative scans through the (111) position
measured on the BT9 spectrometer, where the horizontal bar indicates
the instrumental resolution. (b) Temperature dependence of the
diffuse peak intensities.} \vspace{-4mm}
\end{figure}

In addition, we find a significant temperature-dependence of the
$\vec{Q}$-independent ``background'' signal in
Tb$_{2}$Mo$_{2}$O$_{7}$. Cooling from $T=109$~K to around 50~K, the
intensity at (112) increases roughly linearly with decreasing
temperature.  The $\vec{Q}$-independence of the signal can clearly
be seen in the scans in Fig.1(b) and (c). Typically,
$\vec{Q}$-independent background processes produce scattering which
increases with increasing temperature, which is the opposite of what
we observe.  A likely origin of this $\vec{Q}$-independent
scattering is the presence of frozen Tb moments which are completely
uncorrelated. This identification is supported by the flattening of
the temperature dependence below $\sim 30$~K, which is the same
temperature at which the peaks due to correlated Tb moments begin to
rise quickly. Such $\vec{Q}$-independent scattering from Tb is
unusual and may be caused by their coupling to the slowly
fluctuating Mo moments as discussed further below.  We find that
both the diffuse peaks and the $\vec{Q}$-independent signal are
resolution-limited in energy. Although the energy resolution was
narrow ($\Delta E \simeq 0.12$~meV), fluctuations at a rate below
$\sim 0.02$~THz would be indistinguishable from being static.

Further elastic measurements were performed on BT9, and
representative scans through the (111) peak are shown in the inset
of Fig.~2(a). The solid lines indicate fits to Gaussian lineshapes
plus a sloping background.  The sharp nuclear Bragg peaks, denoted
by the open symbols, were not included in the fits.  Scans were
performed along the $H,H$- and $L$-directions to determine the
static correlation lengths as a function of temperature.
Interpreting the Gaussian linewidths within a finite-size domain
model, we calculated the correlation lengths $\xi$ (or linear domain
sizes) as plotted in Fig.~2(a).  The correlations are not completely
isotropic, with $\xi_{\rm along~[110]}$ slightly larger than
$\xi_{\rm along~[001]}$. Both correlation lengths increase modestly
upon cooling and span a distance of roughly the size of a unit cell.
Again, no dramatic changes are observed at $T_g$. The fitted peak
intensities are plotted in Fig.~2(b). The temperature dependence
closely follows that of the SPINS data (also plotted) even though
the energy resolution for the BT9 measurements is about 5 times
broader. This further confirms that the diffuse peaks are associated
with correlated moments which are static (fluctuating slower than
$\sim$0.02 THz).

\begin{figure}
\centering
\includegraphics[width=8.1cm]{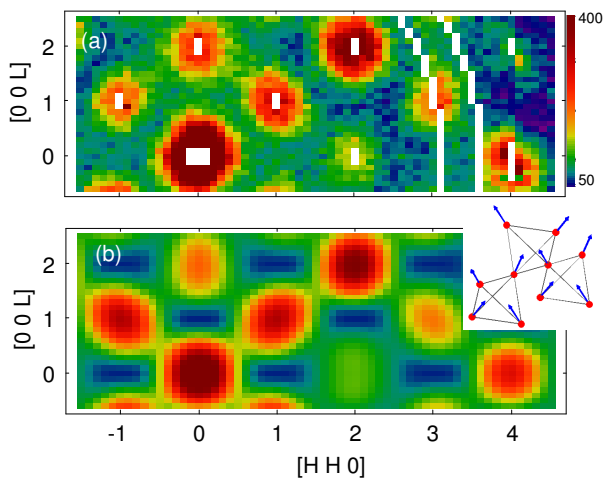} \vspace{-2mm}
\caption{(color online) (a) Pattern of the elastic scattering
intensity at $T=1.5$~K measured on BT9 in the ($HHL$) zone.  Data
points due to scattering from nuclear Bragg peaks and Al powder
lines (indicated by white regions) were removed in order to
emphasize the diffuse magnetic scattering.  (b) Calculated
scattering pattern for short-range ordered regions with the
spin-structure shown in the inset, which are domain-averaged as
described in the text. A constant background of 60 counts (which
approximates the average background in the data) was added to the
calculation.} \vspace{-4mm}
\end{figure}

An important issue is to understand the local structure of the
correlated Tb moments. In Fig.~3(a) we show the pattern of the
elastic scattering intensity in the ($HHL$)-zone taken at $T=1.5$~K
on BT9. In previous powder measurements, two broad peaks were
observed at $|\vec{Q}| \sim 1$~\AA$^{-1}$~and
2~\AA$^{-1}$.\cite{Gaulin_PRL,Apetrei_PRL} Here, the single crystal
data allow for a much more detailed examination of the intensities
throughout reciprocal space.  To first approximation, the peaks
appear at positions consistent with a ``$Q$=0 spin ice'' arrangement
(consisting of equivalent tetrahedra with moments along the local
[111] anisotropy axes with a ``two-in, two-out''
arrangement).\cite{Harris_ice,Gingras_science} However, a closer
inspection of the intensities suggests deviations from the local
spin ice structure.  We calculated the magnetic scattering
cross-section for various spin configurations on a small cluster of
four Tb tetrahedra. A close match was achieved with an arrangement
in which all tetrahedra in the cluster are identical and the Tb
moments are rotated away from the local [111] anisotropy axes of the
spin-ice configuration by $14^\circ$. A depiction of this local
order is shown in the inset of Fig.~3(b).  The rotation of the Tb
moments enhances the net ferromagnetic moment along the
[001]-direction for each tetrahedra. Since there are six equivalent
domains for this spin structure, our calculation averages over all
six magnetic domains.  The uniform susceptibility data rule out a
net ferromagnetic moment at low temperatures, hence the
ferromagnetic moment from each cluster must cancel overall,
consistent with our domain-averaged cross-section. The calculated
cross-section, plotted in Fig.~3(b), shows reasonably good agreement
with the data.  We estimate that the static moment associated with
the diffuse peaks at $T=1.6$~K is $\langle M_{Tb} \rangle \simeq
4.0(5)~\mu_B$, significantly smaller than that expected for a free
Tb$^{3+}$ ion, which may partly be explained by crystal field
effects.\cite{Gingras_PRB,Mirebeau_PRB}  This may also indicate that
a substantial fraction of the Tb moment remains dynamic or is frozen
without measurable spatial correlations. We comment more on the
latter possibility below.  Interestingly, our model for the local
spin structure is similar to the long-range order of Tb moments
observed in diluted Tb$_{1.8}$La$_{0.2}$Mo$_{2}$O$_{7}$ which is a
ferromagnet at low temperatures.\cite{Apetrei_PRL}

Finally, we performed inelastic scattering measurements to
investigate the Tb crystal field excitations, as well as possible
collective excitations.  In the structurally similar
Tb$_{2}$Ti$_{2}$O$_{7}$ and Tb$_{2}$Sn$_{2}$O$_{7}$ compounds, a
low-lying crystal field excitation has been observed near 1.5 meV.
This corresponds to an excitation from the ground state doublet to
the first excited state.  Surprisingly, in Tb$_{2}$Mo$_{2}$O$_{7}$,
we do not observe a sharp mode corresponding to this excitation.
Figure~4(a) shows energy scans taken at the (1.1,\,1.1,\,1) position
on the SPINS spectrometer.  Data taken at the (001) position (not
shown) yield similar spectra.  At $T=1.6$~K, the data for positive
energy transfers reveal a rather flat fluctuation spectrum without a
clear peak in the range between 0.5 meV and 4 meV. Warming to 20 K
produces very little change; though further warming results in
additional low-energy quasielastic scattering, consistent with
previous observations\cite{Gaulin_PRL}. The growth of this
quasielastic scattering appears to correlate with the diminishment
of the $\vec{Q}$-independent elastic signal. Inelastic scans were
also performed at higher energy transfers (up to $\hbar\omega =
20$~meV) on a 2.46 gram powder sample of Tb$_{2}$Mo$_{2}$O$_{7}$
using the BT7 spectrometer, as shown in Fig.~4(b).  No sharp peaks
associated with crystal field excitations are observed; though, the
low temperature spectrum shows broad features centered around 4 meV
and 15 meV. In contrast, both Tb$_{2}$Ti$_{2}$O$_{7}$ and
Tb$_{2}$Sn$_{2}$O$_{7}$ exhibit sharp crystal field excitations at
about 10 meV and 15 meV.\cite{Mirebeau_PRB,Gardner_SL}

\begin{figure}
\centering
\includegraphics[width=7.0cm]{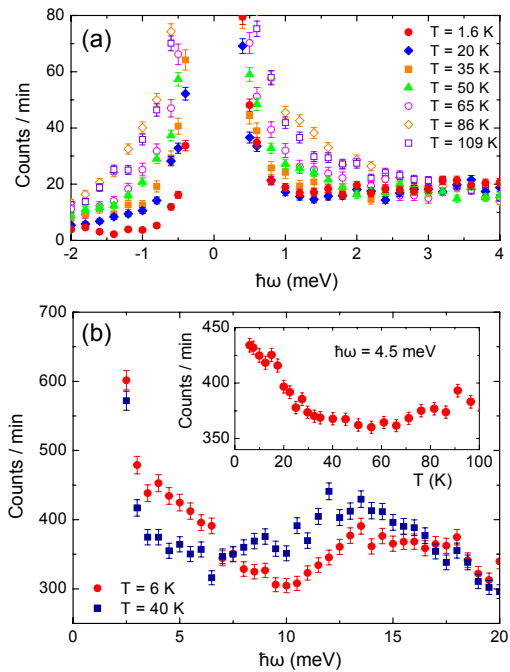} \vspace{-3mm}
\caption{(color online) Inelastic scattering data. (a)
High-resolution measurements taken on SPINS at the (1.1, 1.1, 1)
position in a single crystal.  (b) Higher energy data taken on the
BT7 spectrometer (with $E_f = 14.7$~meV and horizontally focused
analyzer) on a powder sample at $|\vec{Q}|=3$~\AA$^{-1}$. Inset:
Temperature dependence of the scattering at $\hbar\omega=4.5$~meV.}
\vspace{-4mm}
\end{figure}

The lack of sharp crystal field excitations for Tb$^{3+}$ in
Tb$_{2}$Mo$_{2}$O$_{7}$ is unexpected and requires better
understanding. One possible origin may be a static inhomogenous
splitting of the non-Kramers Tb$^{3+}$ doublet due to lattice
strains.  Such an effect has been observed in
Pr$_{2-x}$Bi$_{x}$Ru$_{2}$O$_{7}$ which has obvious chemical
disorder.\cite{vanDuijn_PRL}  While structural disorder has not yet
been identified in Tb$_{2}$Mo$_{2}$O$_{7}$, recent experiments on
the Y$_{2}$Mo$_{2}$O$_{7}$ compound reveal that there exists lattice
disorder associated with the Mo-Mo bond distances\cite{Booth,Keren}.
Since Tb$_{2}$Mo$_{2}$O$_{7}$ has a similar spin-freezing
transition, it is reasonable to expect similar levels of Mo
sublattice disorder. This would produce some degree of inhomogeneity
in the Tb crystal field environments, thereby broadening the
excitations. Another effect to consider is the interaction of the
slowly fluctuating Mo moments with the Tb moments.  Freezing of the
Mo moments manifestly breaks time-reversal symmetry, and therefore
any coupling with the Tb sublattice would split the ground state
Tb$^{3+}$ doublet.  The magnitude of the splitting would depend on
the mean-field generated by the local configuration of Mo moments.
Assuming the Tb$^{3+}$ ground state doublet is primarily $|\pm
4\rangle$ or $|\pm 5\rangle$ as seen in related Tb
pyrochlores\cite{Gingras_PRB,Mirebeau_PRB}, then the mean-field at
the Tb site due to the neighboring Mo would select a preferred local
spin direction. As the Mo moments gradually become frozen in a
disordered arrangement, the net effect on the Tb sublattice would be
to select $|+J_z\rangle$ or $|-J_z\rangle$ in a spatially disordered
way. Such a frozen and uncorrelated Tb configuration would appear as
a $\vec{Q}$-independent background, consistent with our observations
from the elastic scattering measurements.  Indeed, the presence of
static but uncorrelated Tb moments would naturally explain the
relatively small value of $\langle M_{Tb} \rangle$ calculated from
the diffuse peak intensities, since a significant fraction of the
elastic signal would be $\vec{Q}$-independent.

In Fig.~4(b), the inelastic spectrum at $T=6$~K of the powder sample
exhibits enhanced scattering centered around 4 meV which quickly
diminishes upon warming to above $\sim 30$~K (see the inset).  The
temperature dependence closely follows that of the diffuse elastic
peaks. This behavior suggests that the enhanced scattering around
4~meV is derived from collective excitations of the Tb moments
rather than single-ion physics.  The energy for this collective
excitation should thus be related to a combination of the Tb-Tb and
Tb-Mo interaction energy scales. Further inelastic studies on the
single crystal sample would be necessary to specify the magnitudes
of the different interactions.

In conclusion, neutron scattering studies on a single crystal of
Tb$_{2}$Mo$_{2}$O$_{7}$ reveal two components to the elastic
scattering: a set of diffuse peaks plus a $\vec{Q}$-independent
signal. The short-range order of the Tb moments has been identified,
and the observed correlation lengths are slightly anisotropic.  The
inelastic data indicate that the physics of the Tb sublattice is
affected by disorder, most likely through lattice inhomogeneity and
through coupling to the glassy Mo moments. The local clusters of
correlated Tb spins that we observe may be the relevant unit in the
slow dynamics found in this material\cite{Dunsiger} and may connect
to similar physics observed in other Tb-based pyrochlore systems as
well\cite{Bert_PRL,Dalmas_PRL}.

We thank T. Senthil, S. Speakman, and Y. Chen for useful
discussions. The work at MIT was supported by the Department of
Energy (DOE) under Grant No. DE-FG02-07ER46134.  This work used
facilities supported in part by the NSF under Agreement No.
DMR-0454672.

\bibliography{TMO}
\end{document}